\documentclass[english]{bvm2020}

\usepackage{graphicx}
\usepackage{caption}
\usepackage{verbatim}
\usepackage{multirow}
\usepackage{xcolor}
\usepackage{boldline}
\usepackage{float}
\usepackage{ulem}
\usepackage{subfigure}

\newcolumntype{L}[1]{>{\raggedright\let\newline\\\arraybackslash\hspace{1pt}}m{#1}}
\newcolumntype{C}[1]{>{\centering\let\newline\\\arraybackslash\vspace{1pt}}m{#1}}
\newcolumntype{R}[1]{>{\raggedleft\let\newline\\\arraybackslash\hspace{1pt}}m{#1}}

\def\articlenumber{0000}

\date{}
\title{What Do We Really Need?}

\subtitle{Degenerating U-Net on Retinal Vessel Segmentation}

\author{Weilin Fu\inst{1,2} \and Katharina Breininger\inst{1} \and Zhaoya Pan\inst{1} \and
Andreas Maier\inst{1,3}}
\authorrunning{F. Author et al.}
\institute{Pattern Recognition Lab, Friedrich-Alexander Universtiy\and
International Max Planck Research School for Physics of Light (IMPRS-PL)\and
Erlangen Graduate School in Advanced Optical Technologies (SAOT)\\
\email{weilin.fu@fau.de}}
\begin{document}

%==============================================================================
% wählen Sie mit dem Befehl \selectlanguage die Sprache aus, in der Ihr 
% Proceeding verfasst ist
%
%\selectlanguage{german}
\selectlanguage{english}

\maketitle

\begin{abstract}
Retinal vessel segmentation is an essential step for fundus image analysis. With the recent advances of deep learning technologies, many convolutional neural networks have been applied in this field, including the successful U-Net. In this work, we firstly modify the U-Net with functional blocks aiming to pursue higher performance. The absence of the expected performance boost then lead us to dig into the opposite direction of shrinking the U-Net and exploring the extreme conditions such that its segmentation performance is maintained. Experiment series to simplify the network structure, reduce the network size and restrict the training conditions are designed. Results show that for retinal vessel segmentation on DRIVE database, U-Net does not degenerate until surprisingly acute conditions: one level, one filter in convolutional layers, and one training sample. This experimental discovery is both counter-intuitive and worthwhile. Not only are the extremes of the U-Net explored on a well-studied application, but also one intriguing warning is raised for the research methodology which seeks for marginal performance enhancement regardless of the resource cost.
\end{abstract}

\section{Introduction}

Segmentation of retinal vessels is a crucial step in fundus image analysis. It provides information of the distribution, thickness and curvature of the retinal vessels, thus greatly assists early stage diagnosis of circulate system related diseases, such as diabetic retinopathy. Researchers have devoted to this field for decades~\cite{srinidhi2017recent}, and with the development of deep learning technologies \cite{maier2019gentle}, many deep networks have been proposed to tackle this problem. For instance, a Convolutional Neural Network (CNN) combined with conditional random field in~\cite{fu2016retinal}, a network pipeline concatenating a preprocessing net and a vesselness Frangi-Net in~\cite{fu2019divide}, and the U-Net~\cite{ronneberger2015u}.
Since published, U-Net has achieved remarkable performance in various fields. Researchers~\cite{isensee2018nnu} even claim that hyper-parameter tuning the U-Net rather than constructing new CNN architectures is the key to high performance. However, U-Net generally contains huge amounts of parameters and is resource consuming. Previously, researchers~\cite{zhou2018unet++} have proposed to prune U-Net levels in the testing phase to reduce the network size. Yet the modifications introduce even more parameters in the training phase, and only one decisive factor of the architecture, the number of levels, is considered.

We work on retinal vessel segmentation on the DRIVE database and start with a three-level U-Net with 16 filters in the input layer. Firstly, we aim to enhance its performance by integrating common deep learning blocks into the architecture. As the expected performance boost is not observed, we propose the assumption that the basic U-Net alone is adequate or even overqualified for the task. To verify this hypothesis, we design an experiment series to compress the basic U-Net. The number of levels, convolutional layers per level, and filters per convolutional layer are reduced respectively. Non-linear activation layers are removed, and the number of training sets are decreased to further delve into the limits of the network training procedure. Results show that surprisingly harsh conditions are required for the U-Net to degenerate, indicating that the default configuration is redundant. Our contributions are two-fold: the minimum U-Net for this task is reported, indicating the possibility of real-time retinal vessel segmentation on mobile devices; and the issue of excessive computational power use is exposed and stressed on.

\section{Materials and Methods}
\subsection{Default U-Net Configuration}
A three-level U-Net with 16 filters in the input layer is used as the baseline architecture as shown in Fig.~\ref{fig:unet_variants} (a). Batch normalization layers are utilized to stabilize the training process. The deconvolution layers are replaced with upsampling combined with convolution layers to avoid the checkerboard artifact.

\subsection{Additive Variants}
\begin{figure}[h]
    \centering
    \begin{minipage}{0.8\textwidth}
    \subfigure[]
    {\includegraphics[width=\textwidth]{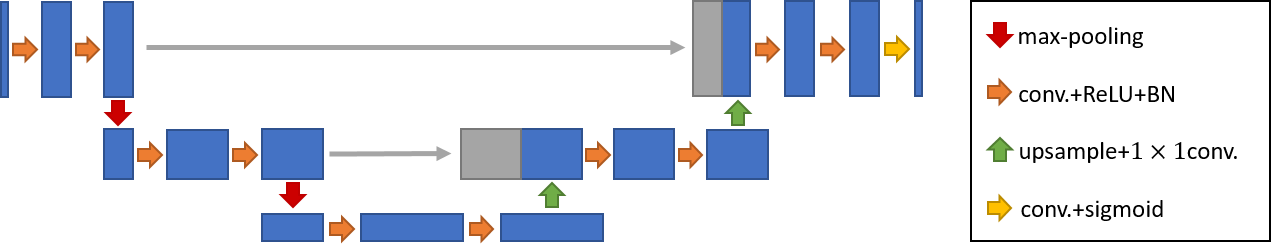}}
    \end{minipage}
    \\
    \begin{minipage}[b]{0.15\textwidth}
    \subfigure[]
    {\includegraphics[height=\textwidth]{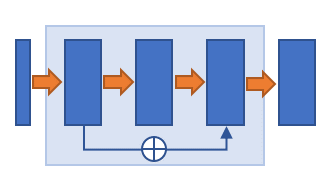}}
    \end{minipage}
    \hspace{2cm}
    \begin{minipage}[b]{0.15\textwidth}
    \subfigure[]
    {\includegraphics[height=\textwidth]{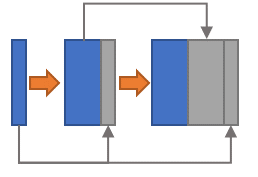}}
    \end{minipage}
    \hspace{1cm}
    \begin{minipage}[b]{0.2\textwidth}
    \subfigure[]
    {\includegraphics[trim=0.5cm 0.2cm 0.2cm 0.5cm, clip, height=\textwidth]{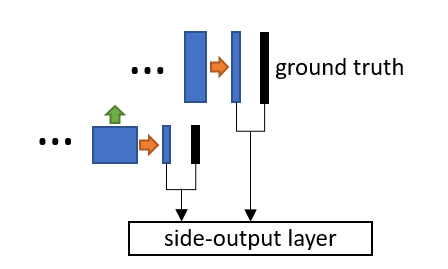}}
    \end{minipage}
    \caption{U-Net (a), residual block (b), dense block (c) and side-output layer (d).}\label{fig:unet_variants}
\end{figure}

Three popular CNN blocks are utilized to modify the network structure. The dense block~\cite{huang2017densely} is inserted in each level of the encoder path. The side-output layer~\cite{fu2016retinal} is employed to provide deep supervision in the decoder path. And the residual block~\cite{he2016deep} is integrated into the encoder, the bottleneck as well as the decoder. The block structures are illustrated in Fig.~\ref{fig:unet_variants} (b-d).

\subsection{Subtractive Variants}
The experiment design of the subtractive variants of the U-Net is based on the ``control variates'' strategy, meaning only one factor is changed from the default configuration in one series. Both the structural and training condition limits of the U-Net are studied in the following experiments:
\begin{enumerate}
    \item The non-linear activation layers, i.e. the ReLU layers, are removed.
    \item The number of convolutional layers in each level decreases to one.
    \item The number of filters in the input layer is reduced from 16 to 1. % by a factor of 2. 
    \item The number of levels decreases step-wise down to one, until the network degenerates into a chain of consecutive convolution layers.
    \item The default U-Net is trained with subsets of the training data. The size of the subset is reduced from 16 down to 1 by a factor of 2.
\end{enumerate}

\subsection{Database Description}
All experiments are trained and evaluated on the Digital Retinal Images for Vessel Extraction (DRIVE) database. DRIVE is composed of 40 RGB fundus photographs with the size of $565\times 584$ pixels. All images are provided with manual labels and Field of View (FOV) masks. The database is equally divided into one training and testing set. A subset containing four images is randomly selected from the training set for validation.

The raw images are prepared with a preprocessing pipeline, where the green channels are extracted, the inhomogeneous illumination is balanced with CLAHE, and pixel intensities within the FOV masks are standardized to (-1, 1). The borders of all FOV masks are eroded by four pixels to remove potential border effects and ensure meaningful comparison. Additionally, multiplicative pixel-wise weight maps $w$ are generated from the manual labels to emphasize on thin vessels using the equation: $w=\frac{1}{\alpha \times d}$, where $d$ represents the vessel diameter in the given manual label, and $\alpha$ is manually set to $0.18$.

\subsection{Experimental Details}
The loss function in this work is composed of two main parts: weighted focal loss~\cite{lin2017focal} and $\ell_2$-norm weight regularizer. The objective function is minimized with the Adam optimizer~\cite{kingma2014adam} with a decaying learning rate initialized to $5\times10^{-5}$. Early stopping is applied according to the validation loss curve. Each batch contains 50 image patches sized $168\times 168$, and data augmentation techniques such as rotation, shearing, additive Gaussian noise and intensity shifting are employed. All experiments are conducted for five times with random initializations to show that the performance is stable and that the conclusion is not dominated by certain specific initialization settings. For subset training experiments, the training sets are selected randomly.

\section{Results}

\begin{figure}[t]
    \centering
    \begin{minipage}{0.28\textwidth}
    \subfigure[Preprocessed Image.]
    {\includegraphics[width=\textwidth]{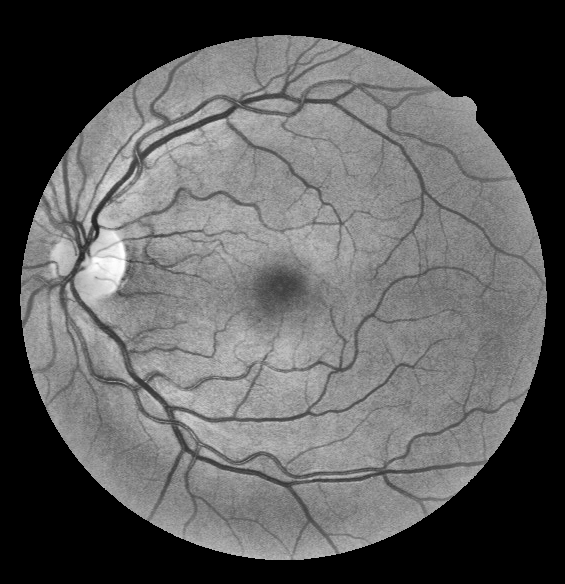}}
    \end{minipage}
    \hfill
    \begin{minipage}{0.28\textwidth}
    \subfigure[Manual label.]
    {\includegraphics[width=\textwidth]{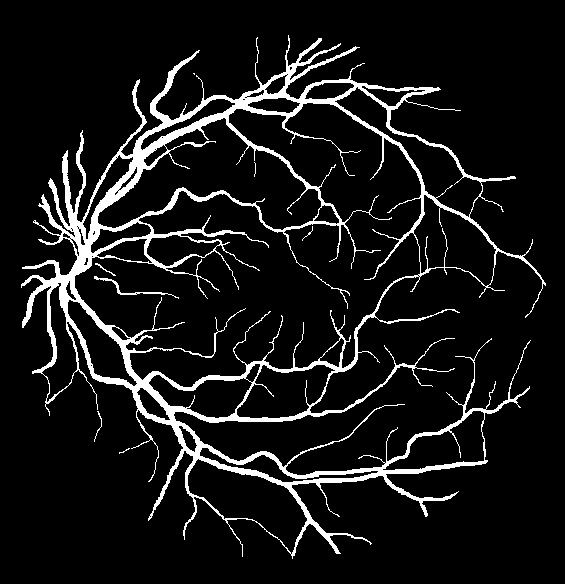}}
    \end{minipage}
    \hfill
    \begin{minipage}{0.28\textwidth}
    \subfigure[Default U-Net.]
    {\includegraphics[width=\textwidth]{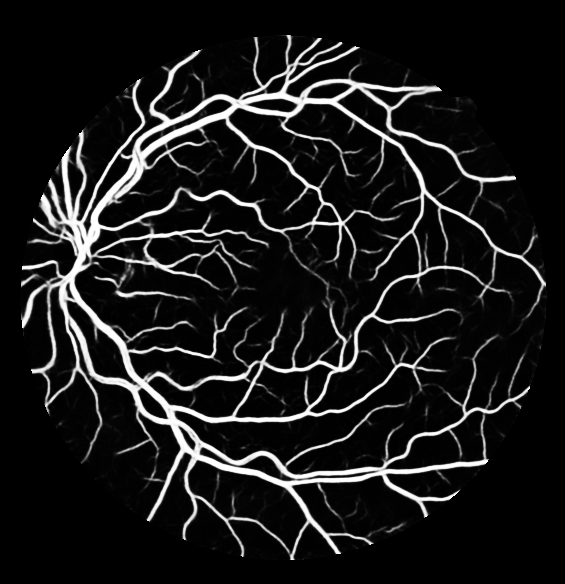}}
    \end{minipage}
    \\
    \begin{minipage}{0.28\textwidth}
    \subfigure[U-Net with 1 filter.]
    {\includegraphics[width=\textwidth]{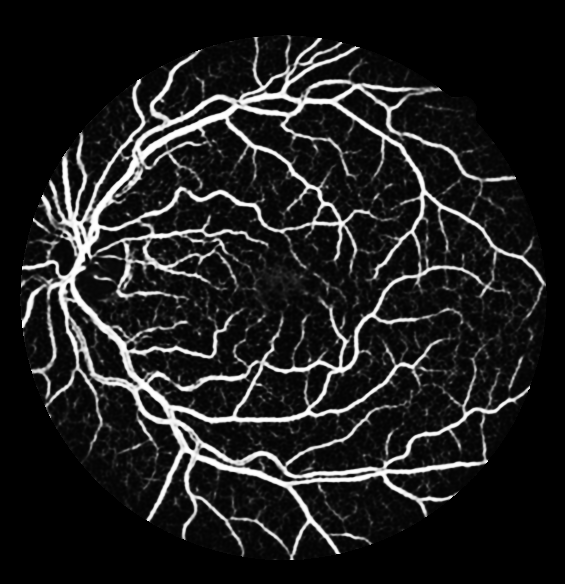}}
    \end{minipage}
    \hfill
    \begin{minipage}{0.28\textwidth}
    \subfigure[U-Net with 1 level.]
    {\includegraphics[width=\textwidth]{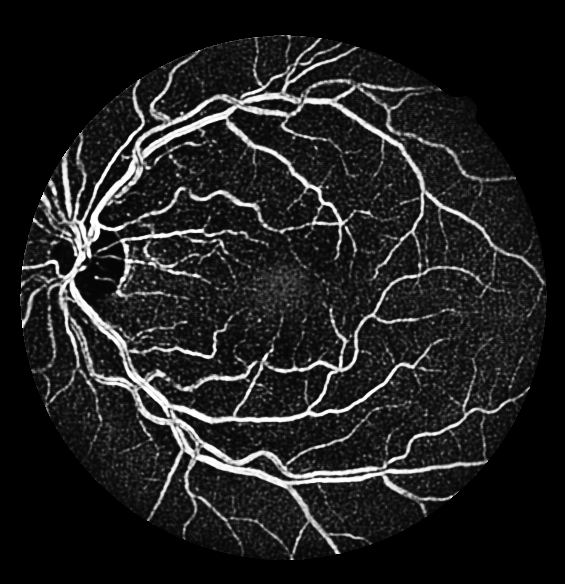}}
    \end{minipage}
    \hfill
    \begin{minipage}{0.28\textwidth}
    \subfigure[Trained with 1 set.]
    {\includegraphics[width=\textwidth]{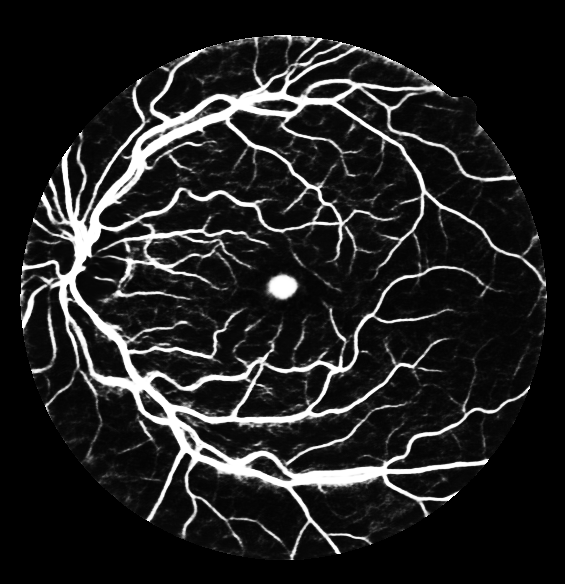}}
    \end{minipage}
    \caption{Probability output of U-Net variants.}\label{fig:result}
\end{figure}

The evaluation of each experiment over the five different initialization roll-outs are reported in Table 1-4. The mean and standard deviations of five commonly used metrics, namely specificity, sensitivity, F1 score, accuracy and the AUC score are presented. The threshold for binarization is selected such that the F1 score is maximized on the validation sets. The threshold independent AUC score is chosen as the main performance indicator. The output probability maps of the degenerated trials are presented in Fig.~\ref{fig:result} (c-f).

Table~\ref{tab:unet_variant_performance} shows that the AUC scores of additive U-Net variants fluctuate merely on the fourth digit, meaning that the expected performance boost is missing. The reduced number of convolutional layers in each level impairs the network marginally, while the absence of non-linearity has an impact on the performance. As for the subtractive experiment series with decreasing numbers of network levels in Table~\ref{tab:u_levelnum_performance} and initial filters in Table~\ref{tab:u_filternum_performance}, surprisingly not until the U-Net contains only one level and collapses into a sequence of convolution layers, or the number of initial filters drops to one, the segmentation results remain satisfactory with an AUC score above 0.97. In respective of the generalization study as reported in Table~\ref{tab:u_setnum_performance}, a monotonous AUC score decline is observed with reducing amount of training subsets, in accordance with our prediction. However we did not anticipate that two sets for training already achieves an AUC score above 0.96, which indicates that the default U-Net has a high generalization capability in retinal vessel segmentation on DRIVE database.

\begin{table}[h]
\centering
\caption{Performance w.r.t. structural variants. Additive variants: Ures, Uden, Uside denote the U-Net with residual blocks, U-Net with dense blocks, U-Net with side-output layers; subtractive variants: U-lin, U-1C represent U-Net without ReLU layers and U-Net with one convolutional layer per level, respectively.}
\label{tab:unet_variant_performance}
\begin{tabular}{L{0.9cm}R{1.5cm}C{2cm}C{2cm}C{2cm}C{2cm}C{2cm}} \clineB{1-7}{2.}
\textbf{var}  & \textbf{param} & \textbf{AUC}  & \textbf{specificity}  & \textbf{sensitivity}  & \textbf{F1 score}  & \textbf{accuracy}   \\ \hline
U              & 108\,976          & $.9748\pm.0005$   & $.9758\pm.0014$           & $.7941\pm.0063$           & $.8101\pm.0010$        & $.9518\pm.0005$         \\
Ures           & 154\,768          & $.9756\pm.0005$   & $.9758\pm.0003$           & $.7994\pm.0033$           & $.8133\pm.0017$        & $.9525\pm.0004$         \\
Uden           & 2\,501\,067         & $.9745\pm.0005$   & $.9742\pm.0013$           & $.8029\pm.0035$           & $.8110\pm.0023$        & $.9515\pm.0008$         \\
Uside          & 109\,072          & $.9744\pm.0006$   & $.9757\pm.0012$           & $.7938\pm.0060$           & $.8097\pm.0023$        & $.9517\pm.0007$         \\ 
\hline
U-lin          & 108\,976          & $.9632\pm.0014$   & $.9693\pm.0022$           & $.7874\pm.0076$           & $.7885\pm.0021$        & $.9453\pm.0010$         \\
U-1C           & 49\,072           & $.9722\pm.0007$   & $.9742\pm.0007$           & $.7918\pm.0044$           & $.8043\pm.0017$        & $.9501\pm.0004$     \\ \clineB{1-7}{2.} 
\end{tabular}
\end{table}

\begin{table}[]
\centering
\caption{U-Net performance w.r.t. different numbers of levels.}
\label{tab:u_levelnum_performance}
\begin{tabular}{C{0.5cm}R{1cm}C{2cm}C{2cm}C{2cm}C{2cm}C{2cm}} \clineB{1-7}{2.}
\textbf{\#} & \multicolumn{1}{c}{\textbf{param}} & \textbf{AUC}   & \textbf{specificity} & \textbf{sensitivity} & \textbf{F1 score}    & \textbf{accuracy}    \\ \hline
2                & 23\,984                               & $.9724\pm.0003$   & $.9733\pm.0016$           & $.7970\pm.0063$           & $.8050\pm.0013$        & $.9500\pm.0007$         \\
1                & 7\,344                                & $.9625\pm.0006$	  & $.9652\pm.0014$           & $.7970\pm.0056$           & $.7832\pm.0015$        & $.9429\pm.0006$  \\
\clineB{1-7}{2.}
\end{tabular}
\end{table}

\begin{table}[]
\centering
\caption{U-Net performance w.r.t. different numbers of initial filters.}
\label{tab:u_filternum_performance}
\begin{tabular}{C{0.5cm}R{1cm}C{2cm}C{2cm}C{2cm}C{2cm}C{2cm}} \clineB{1-7}{2.}
 \textbf{\#}  & \multicolumn{1}{c}{\textbf{param} } & \textbf{AUC}  & \textbf{specificity}  & \textbf{sensitivity}  & \textbf{F1 score}  & \textbf{accuracy}   \\ 
\hline
8                  & 27\,352                               & $.9745\pm .0004$ & $.9754\pm .0007$         &$.7940\pm .0035$         & $.8090\pm .0018$      & $.9514\pm .0005$       \\
4                  & 6\,892                                & $.9739\pm.0003$ & $.9746\pm .0009$         & $.7962\pm .0044$         & $.8080\pm .0010$      & $.9510\pm.0004$       \\
2                  & 1\,750                                & $.9708\pm.0004$ & $.9728\pm .0005$         & $.7889\pm .0028$         & $.7986\pm .0006$      & $.9485\pm .0001$       \\
1                  & 451                                 & $.9620\pm .0010$ & $.9678\pm .0028$	        & $.7776\pm .0102$         & $.7785\pm .0029$      & $.9427\pm .0014$      \\
\clineB{1-7}{2.}
\end{tabular}
\end{table}

\begin{table}[]
\centering
\caption{U-Net performance w.r.t. various number of training sets.}
\label{tab:u_setnum_performance}
\begin{tabular}{C{0.5cm}C{2cm}C{2cm}C{2cm}C{2cm}C{2cm}}
\clineB{1-6}{2.}
\textbf{\#} & \textbf{AUC}   & \textbf{specificity} & \textbf{sensitivity} & \textbf{F1 score}    & \textbf{accuracy}    \\[2pt] \hline
8              & $.9722\pm.0007$   & $.9732\pm.0018$           & $.7961\pm.0093$           & $.8043\pm.0021$        & $.9498\pm.0007$         \\
4              & $.9674\pm.0009$   & $.9700+.0036$           & $.7926\pm.0144$           & $.7935\pm.0013$        & $.9465\pm.0014$         \\
2              & $.9635\pm.0034$   & $.9657\pm.0069$           & $.7919\pm.0180$           & $.7818\pm.0091$        & $.9427\pm.0041$         \\
1              & $.9545\pm.0064$   & $.9672\pm.0049$           & $.7508\pm.0243$           & $.7602\pm.0173$        & $.9387\pm.0047$  \\ \clineB{1-6}{2.}
\end{tabular}
\end{table}

\section{Discussion}
In this work, we explore extreme U-Net configurations for retinal vessel segmentation, and report the results on DRIVE database. This work is motivated by the observation that additive modifications, such as the dense block, introduce additional parameters yet fail to improve the segmentation performance. Hence, an experiment series to decrease the network size as well as simplifying the network structure is conducted. The results do not follow our expectations. It is understandable that non-linearity, rather than the number of convolutional layers per level, has a stronger impact on the network representation capability. However, we did not expect that U-Net with two levels of $23\,984$ parameters, and even U-Net with two initial filters of $1\,750$ parameters can reach an AUC score of over 0.97. Also the generalization ability of U-Net with $108\,976$ weights with only two training sets, achieving an AUC score above 0.96 is surprising. The minimum set-up needed for the U-Net to generate satisfactory results is small for this particular task.

Our discoveries challenge the trend towards networks with increasingly large numbers of parameters that are trained with often marginal improvements in segmentation performance. They also emphasize that, depending on the task, very few samples are sufficient to train CNNs and achieve generalization on unseen data. One can argue that these results are due to the simplicity of the retinal vessel segmentation. Nevertheless, retinal vessel segmentation may not be the only application in this line of observations. We therefore question research approaches merely focused on performance improvement regardless of excessive resource demand. In the future, similar approaches designed under the proposed paradigm could be applied on other tasks to save computational resources.

\section{Acknowledgements}
The research leading to these results has received funding from the European Research Council (ERC) under the European Union’s Horizon 2020 research and innovation programme (ERC grant no. 810316).

\bibliographystyle{bvm2020}

\bibliography{0000}
% Bitte setzen Sie hier Ihre Beitragsnummer ein und benennen Sie
% die BibTeX-Datei ebenfalls auf Ihre Beitragsnummer um.
%Kontrollzeiledef
\marginpar{\color{white}E\articlenumber} % Zeile nicht verändern!
\end{document}